\documentclass[iop]{emulateapj}%[preprint2]{aastex}

\usepackage{graphicx}
\usepackage{epsfig}
\usepackage{epstopdf}
\newcommand{\CaIIIR}{Ca~II~8542}
\newcommand{\Halpha}{H\ensuremath{\alpha}}
\newcommand{\kms}{km~s$^{-1}$}

\begin{document}

\title{Interplay of three kinds of motion in the disk counterpart of type II spicules: up-flow, transversal and torsional motions}

\author{D. H. Sekse\altaffilmark{1}}
\author{L. Rouppe van der Voort\altaffilmark{1}}
\author{B. De Pontieu\altaffilmark{2}}
\author{E. Scullion\altaffilmark{1}}

\affil{\altaffilmark{1}Institute of Theoretical Astrophysics,
  University of Oslo, %
  P.O. Box 1029 Blindern, N-0315 Oslo, Norway}
  
\affil{\altaffilmark{2}Lockheed Martin Solar \& Astrophysics Lab, Org.\ A021S,
  Bldg.\ 252, 3251 Hanover Street Palo Alto, CA~94304 USA}

\begin{abstract}
Recently, it was shown that the complex dynamical behaviour of
spicules has to be interpreted as the result of simultaneous action of
three kinds of motion: (1) field aligned flows, (2) swaying motions,
and (3) torsional motions. We use high-quality observations from the
CRisp Imaging SpectroPolarimeter at the Swedish 1-meter Solar
Telescope to investigate signs of these different kinetic modes in
spicules on the disk. Earlier, rapid blue-shifted excursions (RBEs),
short-lived absorption features in the blue wing of chromospheric
spectral lines, were identified as the disk counterpart of type II
spicules. Here we report the existence of similar absorption features
in the {\it red} wing of the \CaIIIR\ and \Halpha\ lines: rapid {\it red}-shifted
excursions (RREs). RREs are found over the whole solar disk and are
located in the same regions as RBEs: in the vicinity of magnetic field
concentrations. RREs have similar characteristics as RBEs: they have
similar lengths, widths, lifetimes, and average Doppler velocity. The
striking similarity of RREs to RBEs implies that RREs are a
manifestation of the same physical phenomenon and that spicules
harbour motions that can result in a net red-shift when observed
on-disk. We find that RREs are less abundant than RBEs: the RRE/RBE
detection count ratio is about 0.52 at disk center and 0.74 near the
limb.  We interpret the higher number of RBEs and the decreased
imbalance towards the limb as an indication that field-aligned
up-flows have a significant contribution to the net Dopplershift of
the structure.  Most RREs and RBEs are observed in isolation but we
find many examples of parallel and touching RRE/RBE pairs which appear
to be part of the same spicule. We interpret the existence of these
RRE/RBE pairs and the observation that many RREs and RBEs have varying
Dopplershift along their width as signs that torsional motion is an
important characteristic of spicules. The fact that most RBEs and RREs
are observed in isolation agrees with the idea that transversal
swaying motion is another important kinetic mode. We find examples of
transitions from RRE to RBE and vice versa. These transitions
sometimes appear to propagate along the structure with speeds between
18 and 108~\kms\ and can be interpreted as the sign of a transverse
(Alfv{\'e}nic) wave.
\end{abstract}


\section{Introduction}
\label{sec:intro}

The solar chromosphere is a highly dynamical atmospheric layer dominated by numerous small-scale structures.
At the solar limb, some of these small-scale dynamical events can be seen as thin jet-like features protruding from above the photosphere to heights of 10~Mm or more.  
These jets are called spicules and have been known for more than a century, see 
\citet{1968SoPh....3..367B} %Beckers 1968
for a review of the older literature, and 
e.g., \citet{2012SSRv..tmp...65T}  %Tsiropoula et al. 2012
for a more recent review.
\citet{2007PASJ...59S.655D} %DePontieu et al. A tale of two spicules
discovered the existence of a new class of spicules that differed from the known spicules by displaying only upwards motion, shorter lifetimes and higher velocities. 
From an extensive set of data sets, 
\citet{2012ApJ...759...18P}  %Pereira et al. 2012  Quantifying spicules
determined typical velocities for these so-called type II spicules to range between 30--110~\kms, and lifetimes between 50--150~s.
The ``other" type of spicules, or type I spicules, have typical lifetimes of 150--400~s, and show rising and falling phases in parabolic trajectories with maximum rise velocities ranging between 15--40~\kms. 
Type II spicules are the most abundant type which dominate both quiet Sun and coronal hole conditions, only in active regions type I spicules dominate 
\citep{2007PASJ...59S.655D, %DePontieu et al. A tale of two spicules
2012ApJ...759...18P}.  %Pereira et al. 2012  Quantifying spicules
A recent investigation by \citet{2012arXiv1212.2969P} %Pereira+ spatio-temporal effects
demonstrated that only modern observational techniques can provide the required spatio-temporal resolution that is necessary to detect these highly dynamical type II spicules and that bad seeing and technical limitations precluded earlier identification from the ground-based ``classical" observations.

Short lifetimes and high velocities are only two of several factors that contribute to make spicules notoriously difficult to observe: 
vigorous sideways and swaying motion add to its dynamical complexity but the foremost facet
that impedes unambiguous and continuous detection throughout a spicule's lifetime is line-of-sight confusion. 
At and near the limb, the observer's line-of-sight crosses a large number of spicules that combine to a dense forest, in some spectral diagnostics almost a diffuse blanket, 
from which it is nearly impossible to trace all phases of a single spicule's evolution. 
For this reason, the identification of the disk counterpart of type II spicules was regarded as crucial for making progress on the determination of its nature and driving mechanism. 
 \citet{2008ApJ...679L.167L} %Langangen et al. 2008
and
\citet[][Paper~I]{2009ApJ...705..272R} %Rouppe van der Voort et al. 2009 RBE paper I
linked so-called ``rapid blue-shifted excursions" (RBEs), short-lived and spatially elongated spectral asymmetries in the blue wings of the chromospheric \CaIIIR\ and \Halpha\ lines, in on-disk observations to the type II spicules at the limb.
\citet[][Paper~II]{2012ApJ...752..108S} %Sekse et al. 2012 RBE paper II
and 
\citet[][Paper~III]{sekse2013} % Sekse et al 2013 RBE paper III
greatly expanded on the statistics of these events and reinforced the interpretation of RBEs as the disk counterparts of type II spicules.

Besides being one of the main constituents of the chromosphere, interest in type II spicules is motivated by their potential to play an important role in providing mass and energy to the upper solar atmosphere.
Their characteristic sideways swaying motion has been interpreted as a sign that the chromosphere is permeated with Alfv{\'e}nic waves with sufficient energy to heat the quiet Sun corona and accelerate the solar wind
\citep{2007Sci...318.1574D}. %De Pontieu et al. 2007c  Chromospheric alfvenic waves
Studies by \citet{2009ApJ...701L...1D} % De Pontieu et al. roots
and
\citet{2011Sci...331...55D} % De Pontieu et al. Science Hot Plasma 
provided evidence that type II spicules play an important role in mass loading and heating of the corona.
The characteristic rapid fading observed in the {\it Hinode} Ca H observations has been interpreted as a sign of heating of the structure to temperatures that leaves no significant opacity in the observation passband. 

As mentioned above, type II spicules display a complex dynamical behaviour. 
Besides apparent %B
rising velocities that vary over a range of more than 100~\kms, many are displaced 
transversally %B
over distances up to 1~Mm and may reach transverse velocities up to 60~\kms\
\citep{2012ApJ...759...18P}.  %Pereira et al. 2012  Quantifying spicules
Oscillatory swaying motion has been observed with amplitudes of order 10--20~\kms\ and periodicities of 100--500~s 
\citep{2007Sci...318.1574D}. %De Pontieu et al. 2007c  Chromospheric alfvenic waves
\citet{2011ApJ...736L..24O} found a mix of upward and downward propagating as well as standing waves in a large number of spicules in high-cadence {\it Hinode} Ca H limb observations. 
Similar periods and amplitudes were found for \Halpha\ RBEs at disk center in 
Paper~III.
\citet{2008ASPC..397...27S} % Suematsu+ rotation in spicules: double structures
reported that some spicules appear as double and sometimes even multiple threads with evidence of spinning motion. % to add a non-BdP+Oslo paper!
\citet{2012ApJ...755L..11J} % Judge+ evidence -> sudden appearance 
reported the striking observation of spicule-like features that suddenly appear along their full length in extremely high-cadence \Halpha\ wing observations close to the limb. 
Paper~III reported similar behaviour for RBEs at disk center. 
If upward moving plasma were the only motion in spicules, sudden appearance would imply extreme propagation speeds in excess of several thousand \kms. 
%L Judge reports 5000 km/s
However, 
\citet{2012ApJ...752L..12D} %De Pontieu et al. 2012 Twist paper 
established that besides field-aligned flows and swaying motions, a third type of motion, torsional motions of order 25--30~\kms\ are governing the dynamics of spicules. 
When torsional motions are considered, the sudden appearance of spicule-like features in fixed-wavelength imaging can be explained as the result of rotational Dopplershift into the observed passband (Paper~III). 
The observed omni-presence of torsional motion in spicules implies that the combined action of all three modes of motion needs to be considered when interpreting Doppler signals to estimate the transport of plasma and Alfv{\'e}nic wave energy by spicules through the solar atmosphere. 
The ubiquity of torsional motion may also have important implications for determining the production mechanism of spicules and suggest that they may be an important agent for transporting helicity to the upper solar atmosphere. 

In this study, we continue the work by 
\citet{2012ApJ...752L..12D} %De Pontieu et al. 2012 Twist paper 
and Papers~I--III
to investigate signs of torsional motions in the disk counterpart of type II spicules. 
The work of \citet{2012ApJ...752L..12D} %De Pontieu et al. 2012 Twist paper 
was based on various slit and imaging spectroscopic data sets at the limb.
Juxtaposition of the three kinetic modes was observationally established by curved %B
 spectral profiles
in off-limb Ca~II~H $\lambda$-$t$ diagrams, and tilted spectral profiles in $\lambda$-$x$ diagrams of the Ca~II~H, \Halpha, and \CaIIIR\ spectral lines.
In addition, the striking similarity of rows of on-disk and near-limb absorption features in red and blue wing samplings was noted as an important indication of the presence of large-amplitude transversal velocities. 
In the blue wing, these absorption features are comparable to RBEs 
observed closer to disk center. 
Observed in the red wing, these absorption features are more appropriately referred to as ``rapid {\it red}-shifted excursions", or {\it RREs}.  
Here we report the presence of RREs over the whole solar disk
and characterise their properties.

\section{Observations and Data Reduction}
\label{sec:obs}

\begin{table*}[!t]
\begin{center}
\caption{Observational data sets.}
%\begin{tabular}{|l|c|c|c|c|c|c|}
\begin{tabular}{cccccccc}  %no vertical lines!
	\hline \hline
	Data set & Date & Spectral line & \# Line positions & Cadence & Duration & Target & $\mu$ \\ 
	& & & & (s) & (h:m:s) & & \\
	\hline %\hline
	A & 2008 Jun 15 & \CaIIIR\ & 28 & 11.33 & 00:40:25 & CH & 0.987 \\
%	\hline
	B & 2011 May 05 & \Halpha\ & 35 & 7.96 & 00:52:31 & QS & 1 \\
%	\hline
	C & 2011 May 05 & \Halpha\ & 4 & 0.88 & 00:30:53 & QS & 1 \\
%	\hline
	D & 2011 Sep 18 & \CaIIIR\ & 47 & 10.37 & 02:18:53 & QS & 1 \\
%	\hline
	E & 2012 May 26 & \Halpha\ & 33 & (...) & (...) & QS, CH & [1--0] \\
%	\hline
	F & 2012 Jun 02 & \Halpha\ & 33 & (...) & (...) & QS, CH & [1--0] \\
	\hline
\multicolumn{8}{l}{{\bf Note.} --- CH: coronal hole, QS: quiet Sun.} \\
\label{tab:datasets} % inside tabular to get correct labelling....
\end{tabular}
\end{center}
\end{table*}

The observations were obtained at the Swedish 1-m Solar Telescope 
\citep[SST,][]{2003SPIE.4853..341S} %Scharmer et al. 2003a  The swedish 1-m telescope
on La Palma using the CRisp Imaging SpectroPolarimeter 
\citep[CRISP,][]{2008ApJ...689L..69S} %Scharmer et al. 2008  CRISP paper
instrument.  
The CRISP instrument is based on a dual Fabry-P\'{e}rot interferometer (FPI) and contains three high-speed low-noise CCD cameras which operate at a frame rate of 35 frames per second with an exposure time of 17~ms per frame.  All three cameras are located behind the CRISP pre-filter and are synchronised by an optical chopper.  Two of the cameras are also positioned behind the FPI and a polarising beam splitter, while the third camera, which acts as an anchor channel for image processing, is located before the FPI.  
CRISP has a field of view (FOV) of approximately 61\arcsec$\times$61\arcsec and an image scale of 0\farcs0592/pixel.
Fast wavelength tuning ($<$50~ms) is possible between any two positions within the spectral region given by the spectral width of the pre-filter.  With its high speed capabilities CRISP is an ideal instrument for studying the dynamics of the chromosphere through imaging spectroscopy.  Here we are interested in the \CaIIIR\ and \Halpha\ spectral lines.  For \CaIIIR, CRISP has a transmission FWHM of 111~m\AA\ and a pre-filter FWHM of 9.3~\AA.  The transmission FWHM for \Halpha\ is  66~m\AA\ while the pre-filter has a FWHM of 4.9~\AA.  
The combination of the SST adaptive optics system 
\citep{2003SPIE.4853..370S} %Scharmer et al. 2003 Adaptive optics system at the SST
and the Multi-Object, Multi-Frame Blind Deconvolution image restoration technique 
\citep{2005SoPh..228..191V} %MOMFBD paper
results in high spatial resolution, down to the telescope diffraction limit ($\lambda/D$=0\farcs14 for \Halpha). %, of the observations.
Remaining small-scale seeing deformations introduced by the non-simultaneity of the narrowband CRISP images is minimized by using the cross-correlation method developed by 
 \citet{2012A&A...548A.114H}.  %Henriques 2012 - post-MOMFBD destretch

We analyze several high-quality datasets, each with different properties so that various properties of RREs can be addressed. 
Table~\ref{tab:datasets} provides details on the different data sets. 
Data sets A and B are time series with extended wavelength coverage of both wings of the \CaIIIR\ and \Halpha\ spectral lines. 
The other time series of 2011 May 5 (data set C) has an extremely high cadence (0.88~s) so that the dynamical evolution of RREs and RBEs are temporally resolved. Only 4 wavelength positions are sampled, $\pm1032$~m\AA, line center, and the far blue wing at $-2064$~m\AA. %L: check
For more details on the 2011 May 5 data sets, see 
Paper~III. %\citet{sekse2013} % sekse paper III
Data set A was analysed in Paper~I % Rouppe+ RBE paper I
but the data was re-processed for this study: now an efficient correction was applied to deal with artefacts caused by the semi-transparency of the back-illuminated CCD chips at longer wavelengths 
\citep[for details see][]{2012arXiv1204.4448D}. % Jaime PhD thesis
Artefacts were effectively removed and led to a significant increase in the number of RBE detections.

For the other 3 data sets, the observations were recorded under conditions of ``snapshot" seeing: the seeing varied between extremes of total blurriness and crystal-clear sharpness. Most of the time during snapshot seeing, the conditions are useless for science analysis but the intervals of good quality are of sufficiently long duration to cover the acquisition of a complete spectral line scan. 
On 2011 September 18 (data set D), this yielded a large number of disk center \CaIIIR\ line scans with a dense sampling down to 55~m\AA\
(half of the CRISP transmission FWHM). 
On 2012 May 26, \Halpha\ line scans at 7 disk positions were acquired (data set E), starting at disk center and following the central meridian down to the South Pole in intervals of $\mu=0.1$ (with $\mu = \cos(\theta)$ and $\theta$ the observing angle).
On 2012 June 2 (data set F), the central meridian from disk center to the North Pole was covered by 22 %L but we use 21
\Halpha\ line scans (down to 86~m\AA\ sampling), combining into a near-continuous center-to-limb scan (continuous with the exception of one narrow interruption -- see Fig~\ref{fig:ctl_rois}).
Precise pointing information was obtained from co-alignment with full disk images from the 1700~\AA\ channel of the AIA instrument 
\citep{2012SoPh..275...17L} % AIA/SDO
on NASA's Solar Dynamics Observatory.
Using photospheric bright points in the far \Halpha\ wing as reference, sub-arcsecond accuracy in the co-alignment can be achieved.

\section{Method}

For each dataset used in this study, an automated detection routine designed for locating and analyzing RBEs was used to find both RBEs and RREs.  
The same detection method (with slight modifications) %L omit "modifications"?
was used in earlier studies
\citep{2009ApJ...705..272R, % rouppe RBEs Paper I
2012ApJ...752..108S, % sekse Paper II
sekse2013} % sekse Paper III
When searching for RREs, the automated detection routine worked on inverse Doppler images and hence, found RREs in a similar way as RBEs.
For \Halpha, the Dopplergrams were at $\sim$45~\kms, for \CaIIIR\ at 20~\kms.
In the process of detecting RREs we set the threshold of minimum length of the detected events to 20 pixels or $\sim$0.86~Mm (17 pixels for data set A from 2008, when the images scale was slightly larger then for the later data sets). 
This is longer than for our previous studies (17 pixels in Paper~II and 14 pixels in Paper~III) which was found to be necessary in order to avoid detecting false positives.
For example, false positives are caused by the presence of small-scale down-flowing features, or flocculent flows 
\citep{2012ApJ...750...22V}. %Vissers et al. 2012.  Flocculent flow + CRISPEX paper
Flocculent flows are ubiquitously present in the vicinity of magnetic field concentrations and have distinctively different morphology and dynamical behavior than the RREs that we study here.
For each detection, the automated routine records the spatial and temporal position as well as determining the Doppler velocity and Doppler width along the event.  
Doppler velocity and width were determined on a residual profile, obtained from subtracting the RRE/RBE profile from a reference profile that was constructed from averaging line profiles over the whole FOV. %L FOV defined before?
Furthermore, the automated detection routine attempts to link up detected events in successive frames into multi-frame RREs and RBEs which can be used for measurements of lifetimes, transverse motion, and longitudinal motions.

Every dataset was analysed for both RREs and RBEs in order to make comparisons and previously searched datasets were re-examined with the updated length constraint in order to make a proper comparison.

The verification of RREs and the creation of spacetime diagrams was done with the widget based analysis tool CRISPEX 
\citep{2012ApJ...750...22V} %Vissers et al. 2012.  Flocculent flow + CRISPEX paper
which allows for efficient exploration of multi-dimensional datasets.

\section{Results}
\label{sec:results}

\begin{figure}[!th]
\begin{center}
\includegraphics[width=\columnwidth]{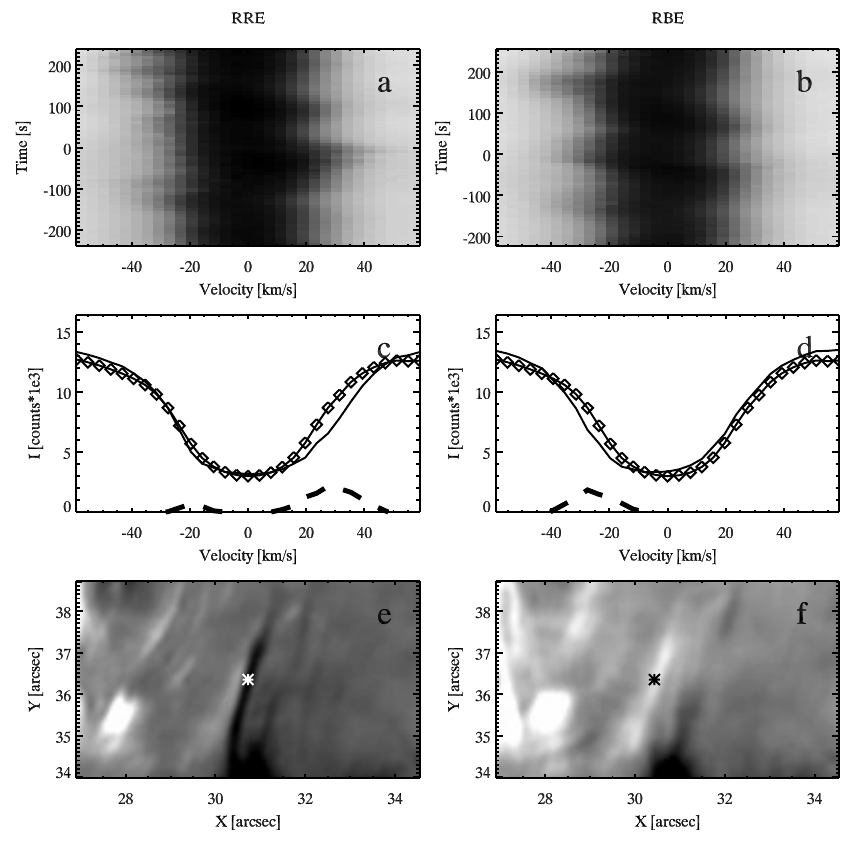}
\caption{Details of an RRE (left column) and an RBE (right column) that occur in rapid succession in neighbouring locations in \Halpha\ (data set B).  The top row shows the temporal evolution of the spectrum at a central location in the RRE and RBE, marked with asterisks in panels (e) and (f).  The middle row shows the detailed spectral profiles (solid line) of the positions marked with asterisks in panel (e) and (f) at the time of the background image.  The line profile marked with diamonds is an average over the whole FOV shown as reference. The thick dashed line is the difference between the average profile and the RBE/RRE profile which is used to calculate the Doppler velocity and width of the event from. The bottom row shows 35~\kms\ Doppler images of the RRE (dark) and RBE (bright) separated by a couple of timesteps.}
\label{fig:ha_details}
\end{center}
\end{figure}
%velocity of the Dopplermaps: 774 mAA or 35.4 km/s
%referee requests colour for dashed line

\begin{figure}[!ht]
\begin{center}
\includegraphics[width=\columnwidth]{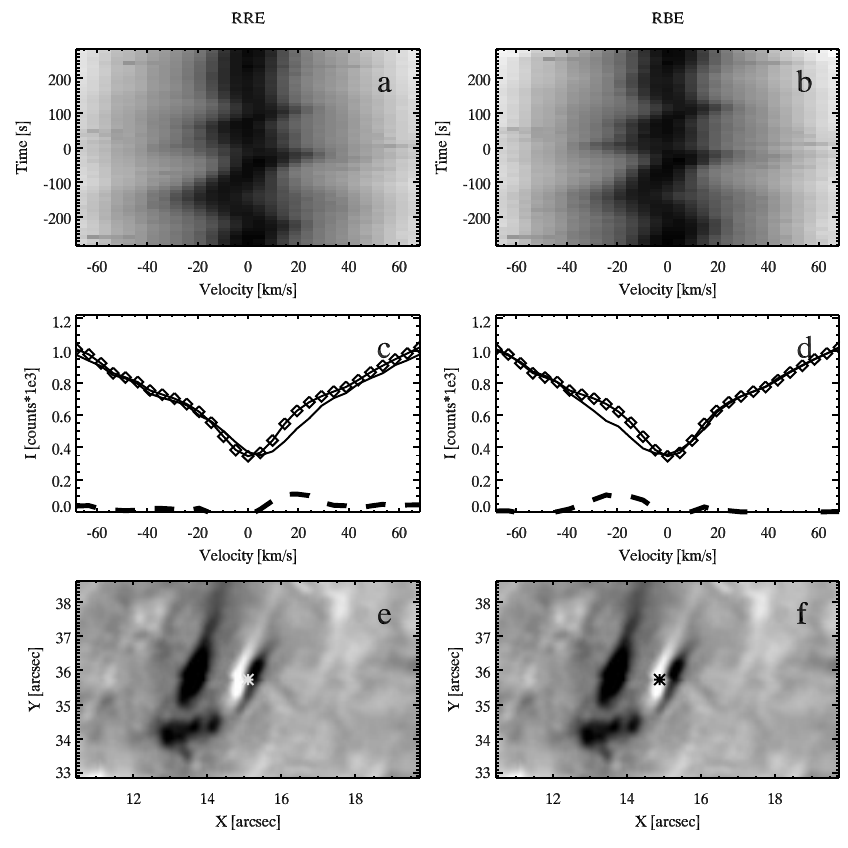}
\caption{Details of an RRE and RBE observed in \CaIIIR\ (data set A). The panels follow the same format as Fig.~\ref{fig:ha_details}. Both the RRE and RBE are present in the same timestep and appear to be part of the same spicule structure. The background 20~\kms\ Dopplermaps in panels (e) and (f) are the same. The same event is shown in more detail in the right column of Fig.~\ref{fig:tran_slit2}}
\label{fig:ca_details}
\end{center}
\end{figure}
%velocity of the Dopplermaps: 582 mAA  or 20.44 km/s
%referee requests colour for dashed line

When searching on-disk \Halpha\ and \CaIIIR\ datasets for excursions in the red wing, many events are detected. 
These events are very similar in appearance and morphology as RBEs.
Figures~\ref{fig:ha_details} and \ref{fig:ca_details} show details of RREs in comparison with RBEs. 
Figure~\ref{fig:ha_details} shows an RRE event observed in \Halpha. Just after the event disappeared in the \Halpha\ red wing, an RBE appeared in the blue wing just adjacent to where the RRE was before. This RBE is shown in the right column of Fig.~\ref{fig:ha_details}.
Figure~\ref{fig:ca_details} shows RRE and RBE events in \CaIIIR. These events were visible as an RRE/RBE pair of adjacent events 
of similar length and morphology
% request from referee: elaborate on impression of being part of same spicule structure. 
giving a strong impression of being part of the same spicule structure. 
The middle panels (c and d) of these figures show the detailed line profiles at a central location in the event, together with reference average line profiles. 
Just like for RBEs, the RREs are characterised by an asymmetry of the line profile with enhanced depression of the line wing -- in the red wing rather than the blue wing. 
Likewise, the temporal evolution for RREs and RBEs is similar: the asymmetry of the line profile is rather short-lived with comparable lifetimes for the two type of events. This is illustrated in the top panels (a and b). 
Furthermore, as can be seen from the bottom panels, the morphology for RREs and RBEs is similar: narrow, elongated structures often protruding from regions with enhanced magnetic fields (bright points, network regions, etc).  

\begin{figure*}[!ht]
\begin{center}
\includegraphics[width=\textwidth]{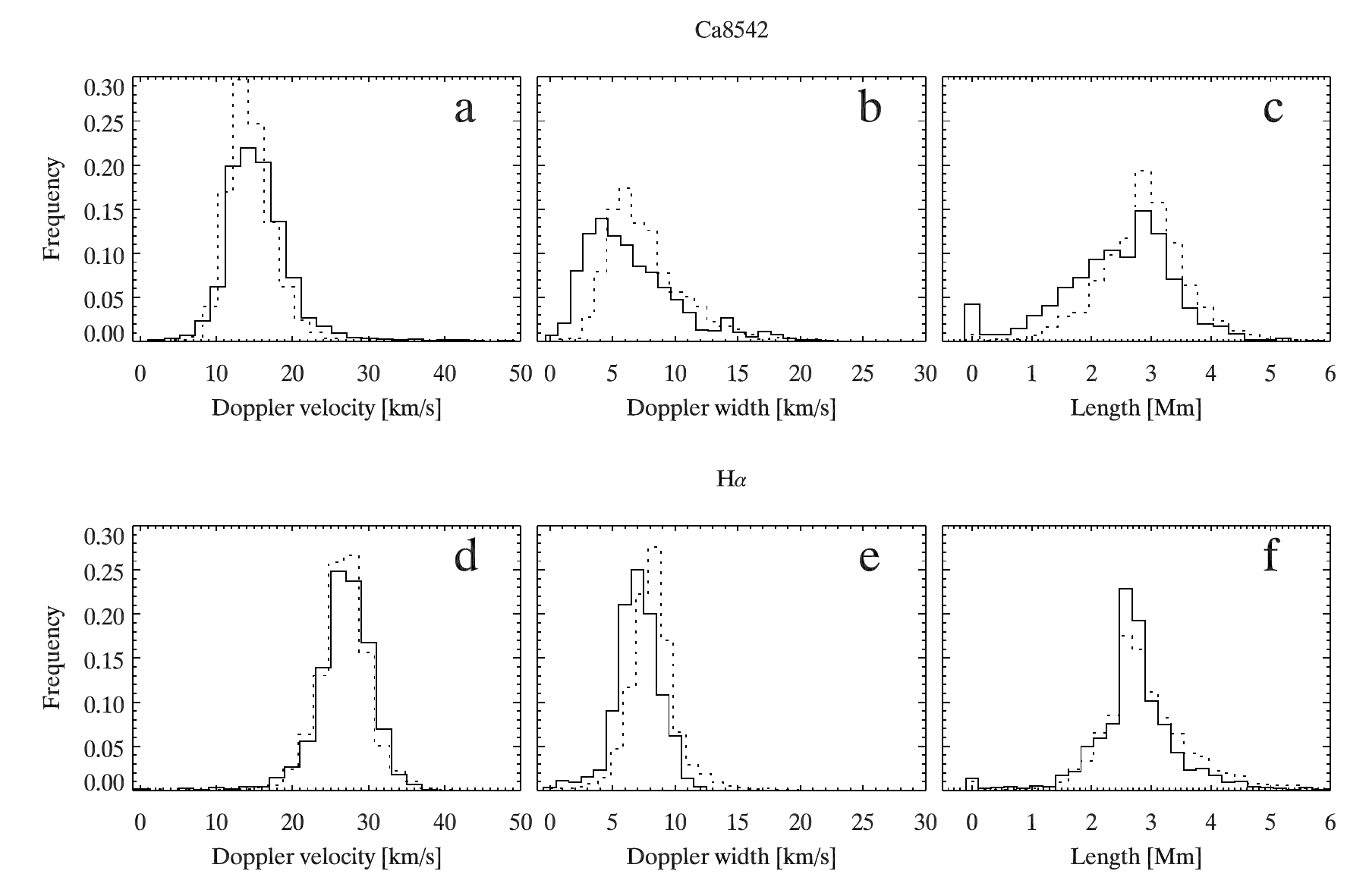}
\caption{Histograms of RRE (solid line) and RBE (dashed line) properties in \CaIIIR\ (top row) and \Halpha\ (bottom row).  Doppler velocities (a and d) and Doppler widths (b and e) are averaged over the length of the event. Lengths are presented in panels (c) and (f). Number of measurements, \CaIIIR\ data set A: 1237 RREs and 2544 RBEs, \Halpha\ data set B: 1354 RREs and 3128 RBEs.}
\label{fig:rre_hist}
\end{center}
\end{figure*}

The measurements of Doppler velocities, Doppler widths, and lengths for %all % not all: excluded singles
detected events in the data sets A (\CaIIIR) and B (\Halpha) are shown in Fig.~\ref{fig:rre_hist}.
As a way to suppress the contribution from possible false detections, we only include measurements on events that are part of a multi-frame event, 
%i.e., events that could be identified over more than one time step. 
i.e., events that could be identified in successive detection images and were therefore connected over a sequence of time steps.
%referee requests more clarity on characterisation time step
%
These measurements were done per time step. 
Doppler velocities and width measurements are averages over the detection length of the event. 
There are significantly more events in the blue wing than in the red wing: 
for the 52~min \Halpha\ quiet Sun time series (B) there are 3128 RBE and 1354 RRE detections (43\%), 
whereas for the 40~min \CaIIIR\ coronal hole time series (A) there are 2544 RBE and 1237 RRE detections.

From the histograms in Fig.~\ref{fig:rre_hist} it is clear that for these properties, RREs (solid lines) are very similar to RBEs. 
Each property shows largely overlapping distributions.
For \CaIIIR\ RREs we see Doppler velocities lying mainly between 10 and 20~\kms, Doppler widths ranging from 1 to 20~\kms, 
and lengths ranging mainly from 1~Mm up to approximately 4.5~Mm.  
These RRE properties are comparable to the RBE properties (dashed lines) found in the same data set with only minor differences in that Doppler widths and lengths for RBEs are shifted to slightly higher values: 
average Doppler width for RREs at 7.0~\kms\ and for RBEs at 8.2~\kms, and  
average length for RREs at 2.5~Mm and for RBEs at 2.9~Mm. 
For \Halpha, the distributions for Doppler velocities (ranging between $\sim$20 and $\sim$35~\kms) and lengths (between $\sim$1.5 and $\sim$4.5~Mm) are almost identical for RREs and RBEs. 
For \Halpha\ Doppler widths, the distribution for RREs is slightly lower: the average for RREs is 7.5~\kms and for RBEs 8.7~\kms.

\begin{figure*}[!ht]
\begin{center}
\includegraphics[width=\textwidth]{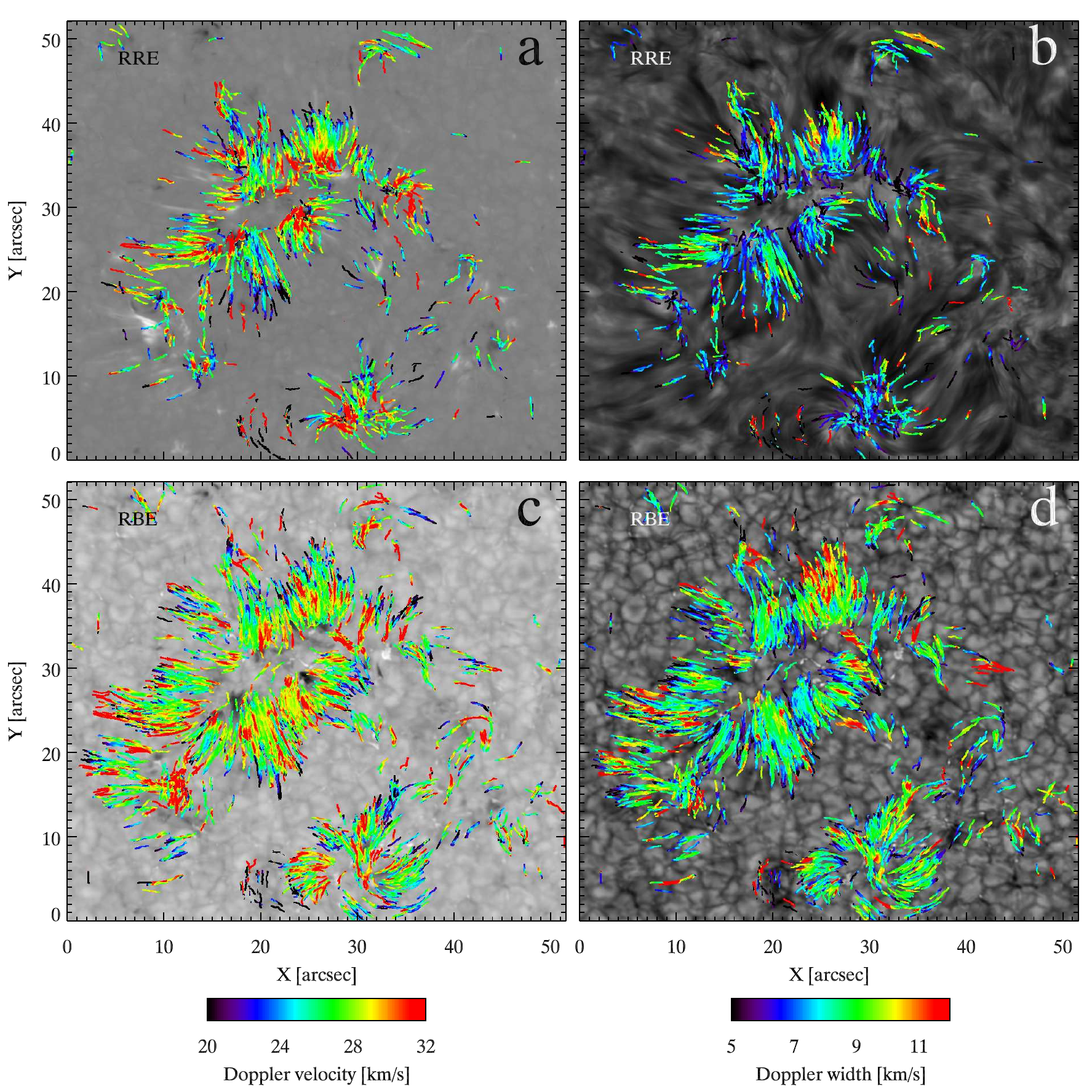}
\caption{\Halpha\ Doppler properties as function of position along RREs and RBEs drawn at their location in the FOV.  RREs are plotted in the top row and RBEs in the bottom row, with Doppler velocities being displayed in the left column and Doppler widths in the right column.  Background images are $\pm$1032~m\AA\ Doppler image (a), line core (b), $+$1032~m\AA\ red wing (c), and far blue wing image (d). }
\label{fig:05may_fov}
\end{center}
\end{figure*}
% All Ha detections are shown: 3994 RBEs, 2084 RREs

\begin{figure*}[!ht]
\begin{center}
\includegraphics[width=\textwidth]{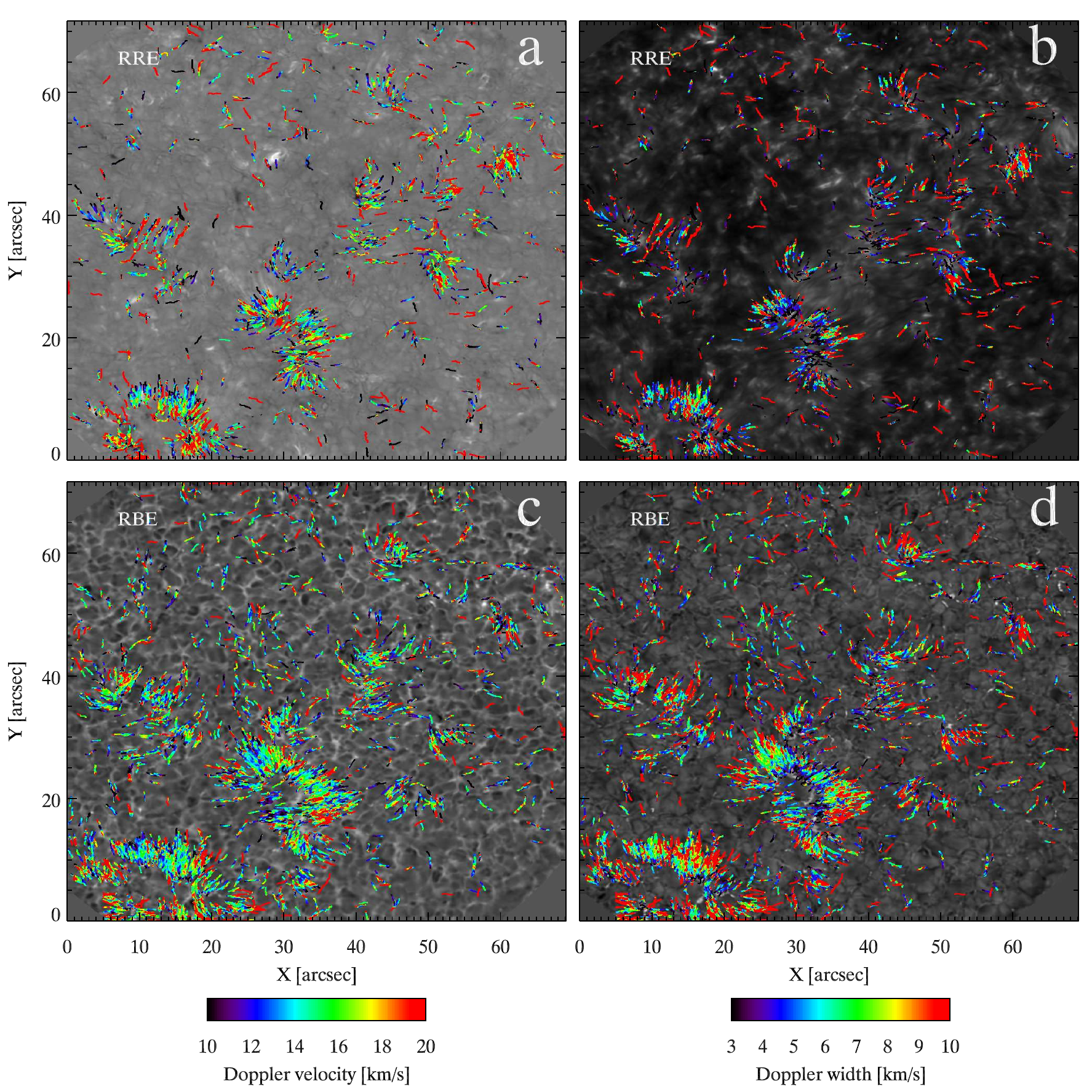}
\caption{\CaIIIR\ Doppler properties as function of position along RREs and RBEs drawn at their location in the FOV. RREs are plotted in the top row and RBEs in the bottom row, with Doppler velocities being displayed in the left column and Doppler widths in the right column. Background images are \CaIIIR\ $\pm$582~m\AA\ Doppler image (a), line core (b), $+$582~m\AA\ red wing (c), and far blue wing (d).}
\label{fig:15jun_fov}
\end{center}
\end{figure*}
%All Ca detections are shown: 
%DL  3655 RBEs  2022 RREs

The variation of Doppler velocity and Doppler width along RREs and RBEs is shown in Figs.~\ref{fig:05may_fov} (\Halpha) and \ref{fig:15jun_fov} (\CaIIIR). 
In these figures, the variation of Doppler signal is shown in colour coding drawn at the physical locations of RREs and RBEs on a background image of the FOV. 
For each pixel, colour indicates the average value for all events that occurred at that location during the time series. 
Comparing the top and bottom rows in these figures, it is evident that there are more events in the blue wing than in the red wing: RBEs are more frequent than RREs (see above for detailed numbers). 
Furthermore, the spatial distribution of RREs is very similar to RBEs: mostly concentrated around regions of enhanced magnetic activity. 
RREs and RBEs occur at the same spatial locations.

Paper~III %\citet{sekse2013} %Paper III
shows comparable maps of \Halpha\ RBE Doppler velocity and width for the same data set (B) in their figure 1 as we do here in Fig.~\ref{fig:05may_fov}c and d. 
The differences are that in Paper~III % \citet{sekse2013} %Paper III
a shorter detection length constraint was used (resulting in 1.7 times more RBE detections) % 6795 RBE detections, 6795/3994 = 1.70
and a slightly different method of drawing these maps: events were drawn in sequential order so that later events mask earlier events. 
The maps in Fig.~\ref{fig:05may_fov}c and d have smoother appearance than their counterparts in Paper~III 
but the same picture emerges: there appears to be no clear general trend for the variation of Doppler properties along the length of RBEs in this disk center quiet Sun time series. 
Similarly, the RRE Doppler properties in the top row panels seem to lack a clear general trend too. 
In the Doppler velocity map of Fig.~\ref{fig:05may_fov}a, there might be a weak
preference for higher velocities towards the footpoints, closer to the magnetic network concentrations (a trend of more red in the lower parts of the RREs).
For the Doppler widths in Fig.~\ref{fig:05may_fov}b, there might be %seems to be %L tone down 
a preference for lower widths towards the footpoints (more blue towards the bottom). 
Note that the trend of more dark and blue colours in the RRE Doppler width map as compared to the RBE Doppler width map reflects the lower Doppler width distribution for RREs in Fig.~\ref{fig:rre_hist}e. 

The Doppler property maps for \CaIIIR\ RBEs in Fig.~\ref{fig:15jun_fov}c and d can be compared with the left panels of figure 12 of \citet{2009ApJ...705..272R}. %Rouppe 2009 RBEs Paper I
They showed Doppler properties of 413 \CaIIIR\ RBEs from the same data set but detected in Dopplermaps at 30~\kms.
Here we show results from 3655 \CaIIIR\ RBE events detected in 20~\kms\ Dopplermaps. 
We confirm the results from Paper~I: in this coronal hole data set, there is a trend for increasing Doppler velocity and width along RBEs (in the colour scheme of this figure: a trend of more red towards the top end of RBEs). 
For \CaIIIR\ RREs in the top panels of Fig.~\ref{fig:15jun_fov} the variation is much more diverse and it is hard to see any clear trend. 

\subsection{Center to Limb Variation}
\label{sec:center2limb}

\begin{figure}[!ht]
\begin{center}
\includegraphics[width=\columnwidth]{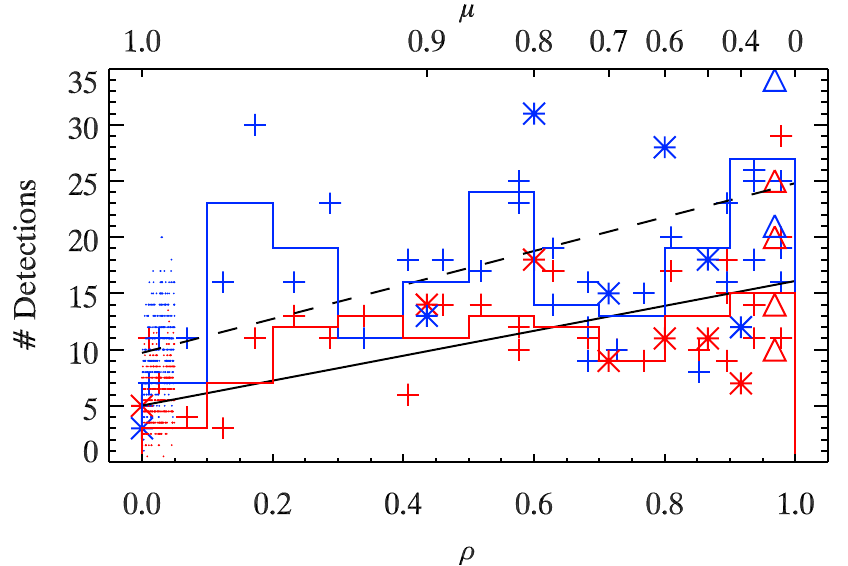}
\caption{Number of detections of RREs (red) and RBEs (blue) per time step over the FOV of several datasets going from disc center ($\rho=r/R_\odot=0$) to the solar limb ($\rho=1$).  Corresponding $\mu=\cos({\rm observing \,\, angle})$ scaling is shown on top. Dots near $\rho=0$ signify data points from the disk center data set B, the spread in $\rho$ is artificial to show the density of data points. Stars are from data set E, crosses are from F. The histogram bars are averages over the data points in $\Delta \rho=0.1$ intervals. Linear fits to the data points are shown with a solid (RREs) and dashed (RBEs) line.}
\label{fig:ctl_detection}
\end{center}
\end{figure}

To further explore the occurrence of RREs on the solar disk, we ran the detection method on data sets at various disk positions. 
The detection was done on Dopplergrams at about the same wavelength offset for all different data sets: approximately $\pm45$~\kms.
Figure~\ref{fig:ctl_detection} shows the number of RRE and RBE detections as function of disk position $\rho = r / R_\odot$ with $r$ the distance from disk center and $R_\odot$ the solar radius. 
The numerous dots around $\rho=0$ are the detection numbers from every frame of the 52 min data set A taken at disk center (396 time steps).
Generally, more RBEs are detected than RREs.
Although there is a significant scatter to the number of detections, we see a clear trend of increasing numbers of both RREs and RBEs being detected towards the limb.
The solid and dashed lines are linear fits to the RRE and RBE detection numbers, respectively, and show an increase in the average number of detections from 5.3, at disk center, to 14.4, near the limb, for RREs, and from 10.1 to 19.4 for RBEs.

\begin{figure*}[!ht]
\begin{center}
\includegraphics[width=\textwidth]{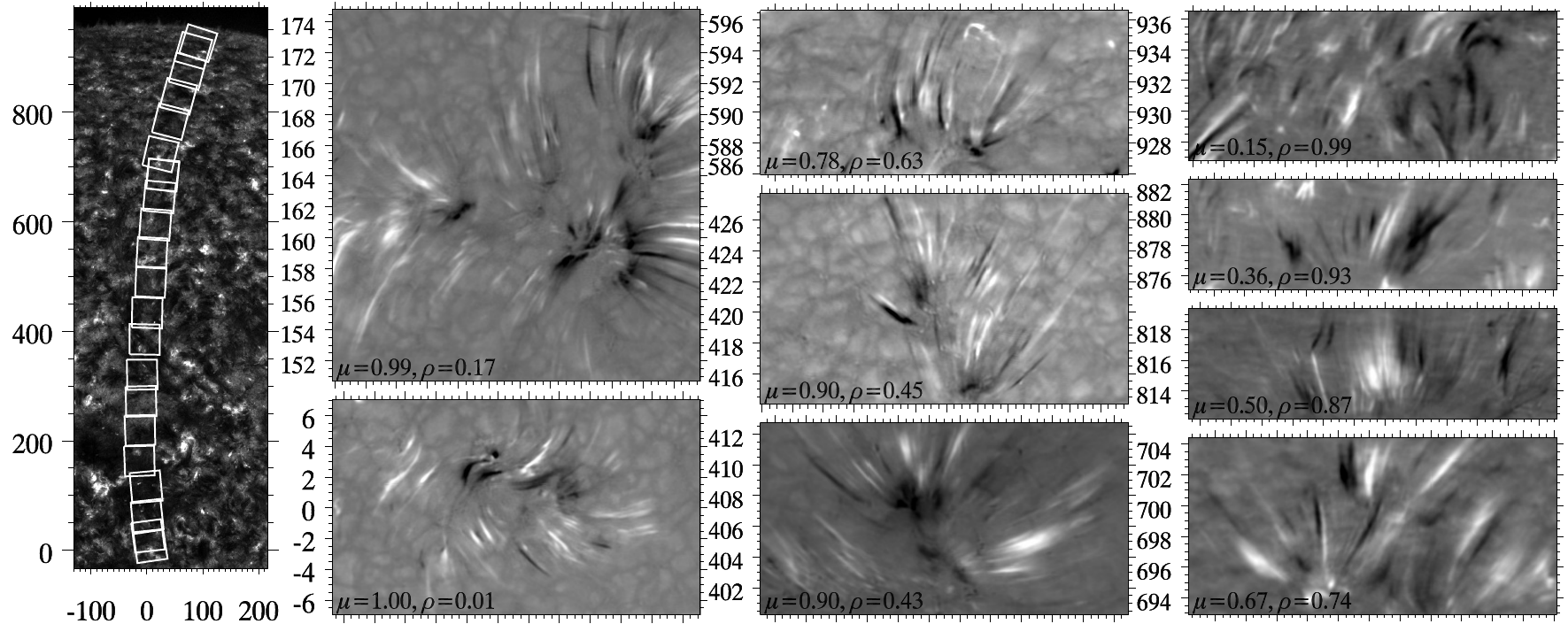}
\caption{Examples of RRE + RBE concentrations around network regions at different disk positions from the center-to-limb scan of data set F. The left panel shows the various telescope pointings as white squares on the background of a co-temporal SDO/AIA 304~\AA\ image. The 9 images are cutouts from $\pm1032$ \Halpha\ Dopplergrams, each centered on network concentrations with enhanced RRE/RBE activity. RREs are white, RBEs are black. Axis labels in heliocentric coordinates (arcseconds).}
\label{fig:ctl_rois}
\end{center}
\end{figure*}
% center-to-limb scan: mostly quiet Sun, crossing a large coronal hole in  the range y=0-210. (defined in SDO/AIA 193 AA)

Figure~\ref{fig:ctl_rois} shows a number of close-up images centered on network regions at various disk positions. 
The center-to-limb scan of data set F mainly covered quiet Sun, it crossed a large equatorial coronal hole between heliocentric $y$=0\arcsec--210\arcsec. 
The upper-left close-up image is from the coronal hole ($y$=152\arcsec--174\arcsec).
Over the whole disk, RREs and RBEs are observed in close vicinity. 
Most often, RREs and RBEs are seen in isolation, but there are a number of examples of RRE/RBE pairs where the RRE and RBE signal is observed in parallel and touching structures, suggestive of being part of the same physical entity. 
A prominent example is visible in the lower left panel ($y$=694\arcsec--704\arcsec) but close inspection reveals examples in every close-up image.

\subsection{Torsional motion}  
\label{sec:transverse_slit}

\begin{figure}[htbp]
\begin{center}
\includegraphics[width=.49\columnwidth]{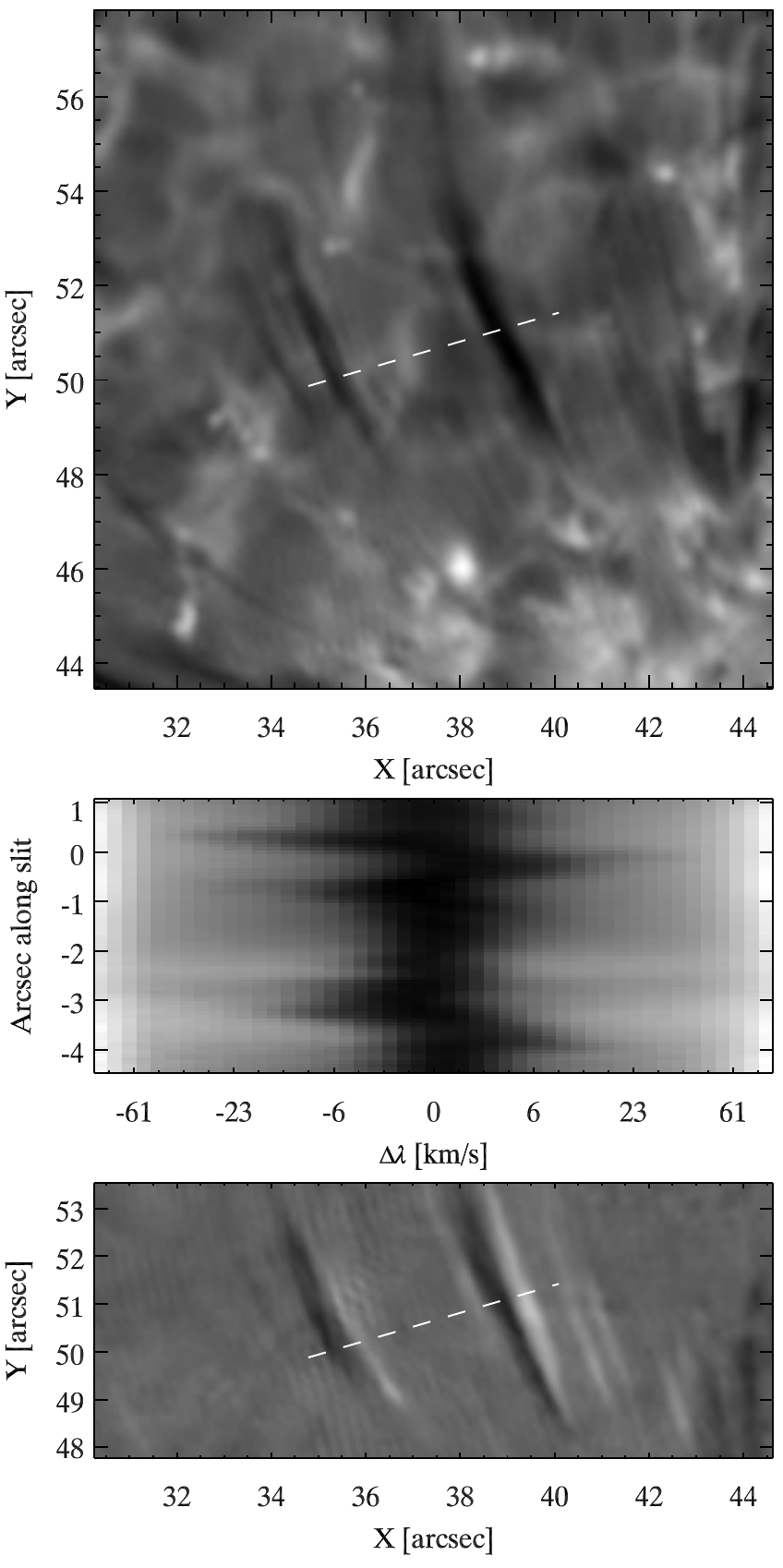}
\includegraphics[width=.49\columnwidth]{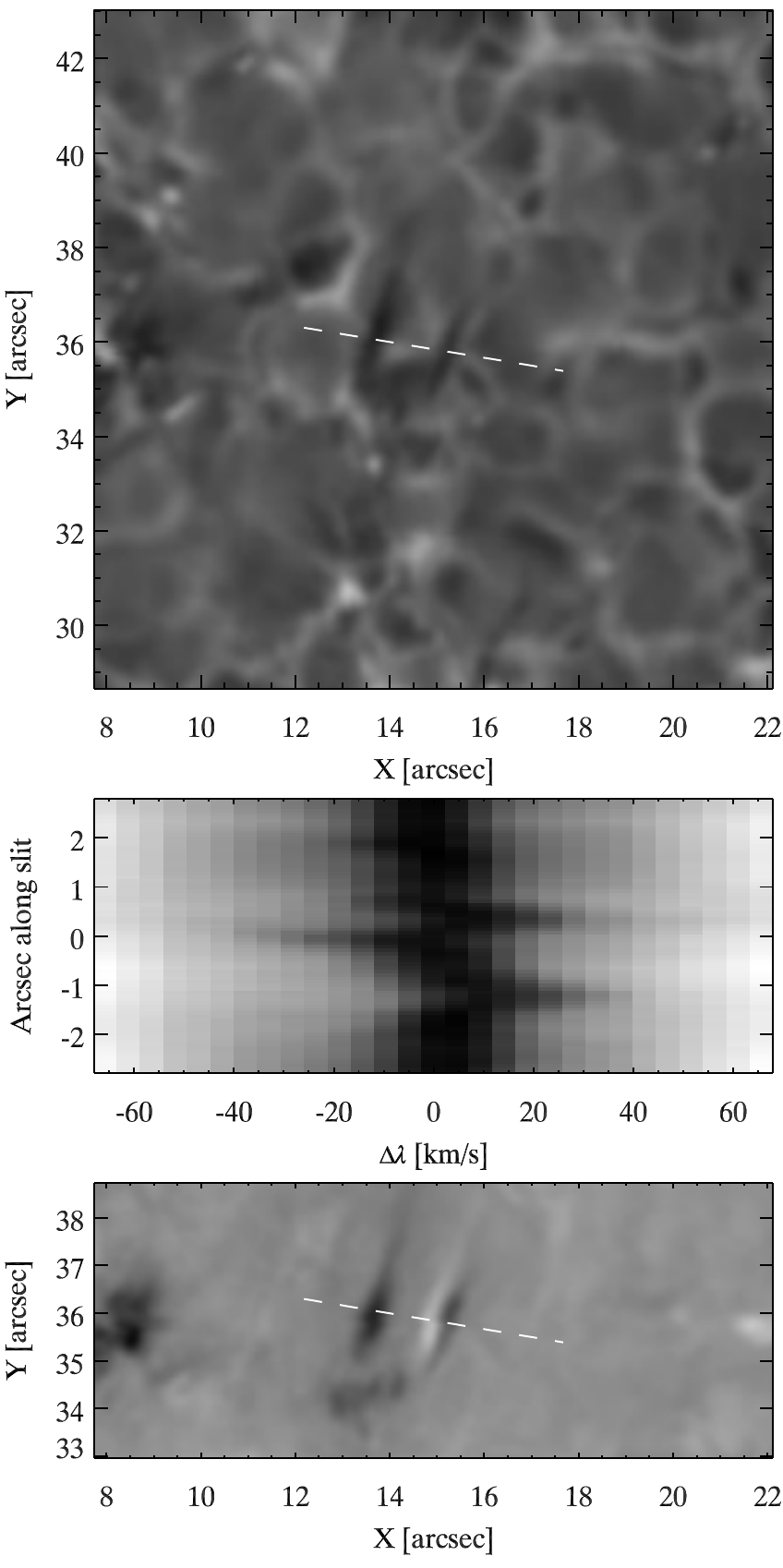}
\caption{Examples of direct evidence of torsional motion in the disk counterpart of spicules. Spectral profiles are extracted from \CaIIIR\ datasets (data set D in left column, data set A in the right column) in a direction perpendicular to the main axes of a number of RREs and RBEs (the "slit" drawn as white dashed lines in the \CaIIIR\ images in the top panels). Top panels show \CaIIIR\ red wing images (left: $+440$~m\AA, right: $+582$~m\AA). Corresponding spectrograms (or $\lambda$-$x$ diagrams) are shown in the middle panels, with the origin of the y-axis centered on the main RRE/RBE pair. The bottom panels show $\pm16$~\kms\ (left) and $\pm20$~\kms\ (right) Dopplergrams with RREs black and RBEs white. Figure~\ref{fig:ca_details} shows more details of the right hand-side event.}
\label{fig:tran_slit2}
\end{center}
\end{figure}

Parallel and touching RRE/RBE pairs are found over the whole solar disk (see Sect.~\ref{sec:center2limb}). 
Figure~\ref{fig:tran_slit2} shows examples from high spectral resolution \CaIIIR\ scans close to disk center.
The middle panels show spectrograms ($\lambda$-$x$ diagrams) from an artificial ``spectrograph slit" drawn in the direction transversal to the main axes of selected events close to a network region. 
The asymmetry of the spectral profiles changes rapidly from one wing to the other over short distances, resulting in tilted profiles in the spectrograms.
For the main event in the left column, centered on 0\arcsec\ along the slit in the middle panel, the asymmetry changes from approximately $+20$~\kms\ to $-20$~\kms\ over only ~250~km. 
Another less extreme example is visible at $-3.5\arcsec$ along the slit. 
The right column shows an example of an RRE/RBE pair with the asymmetry varying between approximately $\pm20$~\kms\ over the width of the structure.

\subsection{Temporal change between RREs and RBEs}
\label{sec:longitudinal_slit}

\begin{figure*}[!t]
\begin{center}
\includegraphics[width=\textwidth]{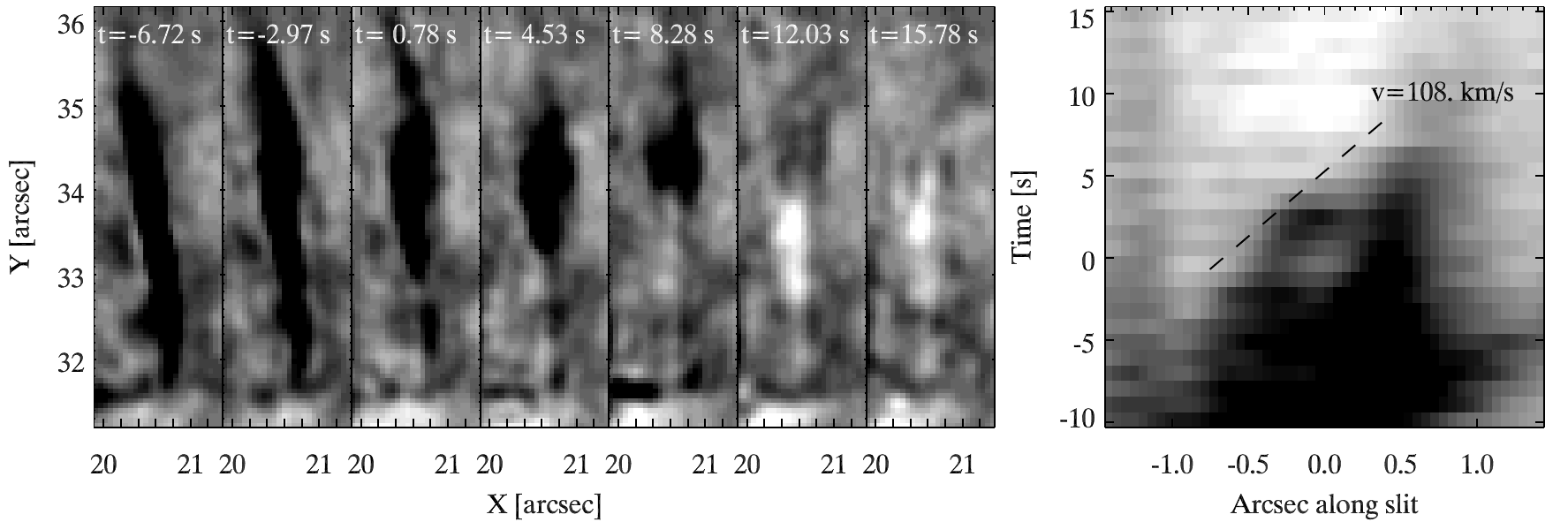}
\includegraphics[width=\textwidth]{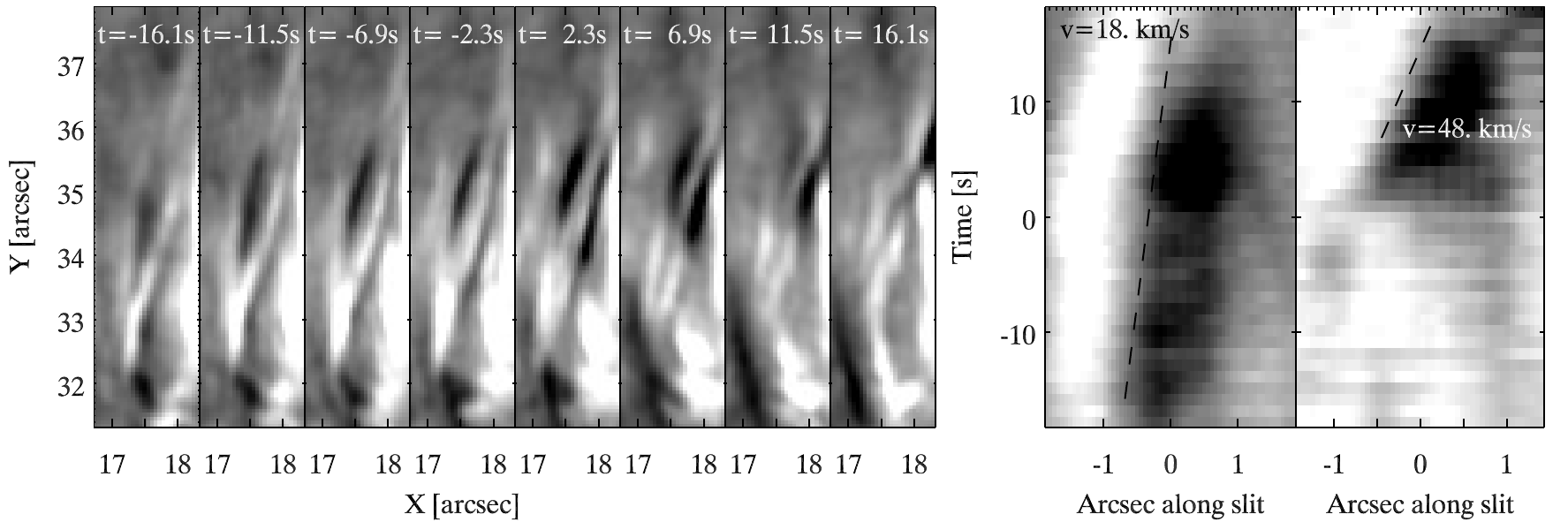}
\caption{Examples of the rapid change from RRE to RBE in the  fast \Halpha\ dataset C. Left panels: sequential images showing the time evolution of the transition from one form into the other.  The right panels show the space-time ($x$-$t$) diagrams from a slit drawn along the RRE/RBE axis. The dashed lines serve as reference for a propagation speed for the transition. In the bottom panels, two examples of transitions from RRE to RBE can be seen. Both events are diagonally oriented in the images on the left. The left $x$-$t$ diagram is associated with the RRE (dark feature) that is visible from the start of the sequence of images. The right $x$-$t$ diagram is associated with the event slightly below that does not appear as an RRE before $t=0$.}
\label{fig:long_slit}
\end{center}
\end{figure*}

Besides parallel and touching RRE/RBE pairs, we frequently observe events for which the asymmetry changes from red-ward to blue-ward and vice versa. 
These changes are very rapid and the most convincing examples that these modulations really are occurring in the same physical structure can be found in the high temporal resolution dataset C (cadence 0.88~s).
Figure~\ref{fig:long_slit} shows a few examples.
The top panels show an event that starts as a long RRE of about 3\arcsec\ length which changes into a short RBE that appears to be growing from the bottom. 
While the RBE appears, the RRE vanishes. 
This change occurs at a propagation speed that is estimated to be about 108~\kms.
The bottom panels show two events that appear close to each other. 
The first event starts as an RRE (dark feature) that is centered on $\sim(17\farcs5,34\farcs5)$ at $t=-16.1$~s and moves slowly (diagonally) upwards. 
At the same time, an RBE slowly grows from the bottom with a slight acceleration after about $t=5$~s. The average propagation speed of the change is approximately 18~\kms. 
Almost parallel to this event, there is a similar event that starts as an RRE that is not visible before $t=0$~s, the space-time diagram is shown as the right-most panel. 
While the RRE slowly moves upward, an RBE grows at an approximate speed of 48~\kms. 
Similar space-time diagrams as the right panels of Fig.~\ref{fig:long_slit} were made for 7 
other examples, finding propagation speeds between 18
and 102~\kms. 
The average over all events is 58~\kms.

\section{Discussion}
\label{sec:discussion}

The red wing counterpart of RBEs: narrow, elongated and short-lived absorption features in the \CaIIIR\ and \Halpha\ lines in the form of asymmetries in the {\it red} wing, or RREs, can be observed in large abundance over the whole solar disk. 
They are very similar to RBEs in terms of length and average Doppler velocity, while the average Doppler width is slightly lower for RREs than for RBEs. 
Their spatial distribution is similar to RBEs: they are found around concentrations of enhanced magnetic field. 
We did not do a detailed study of the temporal evolution of RREs like in Paper~III, 
but cursory inspection of our data suggests that the lifetimes of RREs are similar to RBEs. 
RREs are less abundant than RBEs, around disk center about half the number density, but significant numbers can be found anywhere on the disk.

When we reflect on our own work on the disk counterparts of type II spicules, we can identify several reasons why RREs were not identified earlier.
One reason is that the observations from the early CRISP campaigns were not optimally suited to identify RREs.
Limiting factors for the two data sets from 2008 that were used in Paper~I: the \Halpha\ time series had an asymmetric spectral sampling that did not cover the red wing further than 800~m\AA. 
The \CaIIIR\ time series (data set A in this study) suffered from artefacts related to the semi-transparency of the CCD chip at long wavelengths 
\citep[see][]{2012arXiv1204.4448D}. % Jaime PhD thesis
For this data set, this problem was not resolved until 2011. 
The 2010 observations acquired for follow-up studies (used in Paper~II) were specifically designed for RBE detection: the \Halpha\ line was only sampled in the blue wing, and the \CaIIIR\ sampling was relatively coarse. 

It was not until high resolution off-limb Ca~II~H slit spectroscopy could be combined with high quality CRISP limb observations that the ubiquitous presence of torsional motions in spicules could be unequivocally established
\citep{2012ApJ...752L..12D}. %De Pontieu et al. 2012 Twist paper 
It was realised that spicules undergo three different types of motions at the same time: (1) field-aligned flows of order 50--100~\kms, (2) swaying motions of order 15--20~\kms, and (3) torsional motions of order 25--30~\kms. 
The presence of these three kinetic modes explained the appearance of spicular features in near-limb blue and red wing CRISP \Halpha\ and \CaIIIR\ images.
Bushes of red-shifted and blue-shifted features appear with similar morphology and orientation. 
Field-aligned flows towards and away from the observer alone cannot explain the striking similarity as the asymmetry of near-limb line-of-sight projection should cause large morphological inequality. 
Doppler modulation due to transverse motions from modes (2) and (3) can explain the presence of near-cospatial and morphologically similar absorption features in the blue and red wings of the near-limb CRISP observations. 
This understanding led us to this work investigating the presence of similar red-wing absorption features at disk positions further away from the limb. 

We interpret the presence of RREs close to disk center as a strong indication that field-aligned flow cannot be the only type of plasma motion in spicules. 
High-amplitude transversal motions must be present too in order to produce an effective Dopplershift into the red wing.
We find evidence for these transversal motions to be in the form of torsional motions and/or in the form of swaying motions.
 
 \begin{figure}[!t]
\begin{center}
\includegraphics[width=\columnwidth]{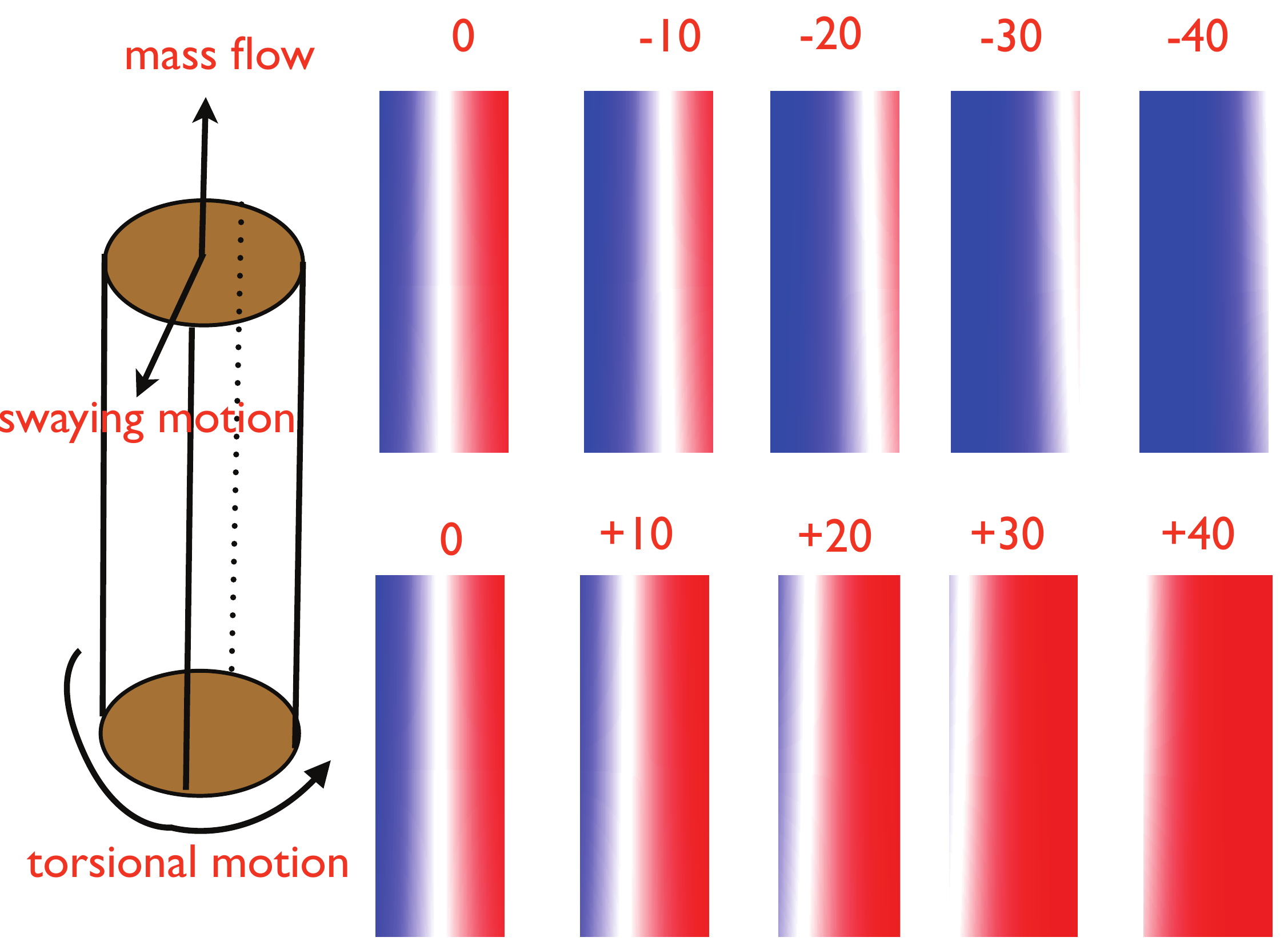}
\caption{Illustration of the effect of the line-of-sight component from up-flow and/or swaying motion to the Doppler profile of a tube that undergoes torsional motion. The tube geometry is sketched on the left-hand side, with the observing angle $\ne 90^\circ$ to the tube's long axis so that both the up-flow along the tube and the swaying motion can have a non-zero component along the line-of-sight. The two rows with panels on the right show the Doppler profile in blue/red shift for varying line-of-sight velocities (denoted in \kms\ at the top, for a 40~\kms\ line-of-sight component from the torsional motion).}
\label{fig:sketch}
\end{center}
\end{figure}

The observation of parallel and touching RRE/RBE pairs is a direct indication of torsional or spinning motion in spicules (see Figs.~\ref{fig:ca_details}, \ref{fig:ctl_detection}, and \ref{fig:tran_slit2}). 
We find a significant number of RRE/RBE pairs in our data but most common are isolated RREs and RBEs. 
The spectral profiles in $\lambda$-$x$ diagrams extracted perpendicular to the main axes of isolated events in high-spectral resolution observations often show signs of tilts similar (but less pronounced) to the extreme cases of the RRE/RBE pairs in the middle row of panels in Fig.~\ref{fig:tran_slit2}. 
This is most clear in the \CaIIIR\ line which has a narrower core than \Halpha\ and therefore more pronounced Doppler modulation. 
This suggests that torsional motion is likely to be present in a large fraction of spicules and therefore an important kinetic mode. 

The importance of swaying as a kinetic mode in RBEs was already evident from transverse motion measurements (see Papers~I--III). 
The observation that most events are either RRE or RBE and not pairs can be interpreted as swaying producing significant offsets in the Dopplershift of these structures. 
Field-aligned up-flows can also result in a Doppler offset so that the dominance of isolated RRE and RBE events is probably due to the combined effect of both kinetic modes. 
Figure~\ref{fig:sketch} illustrates the effect of swaying and/or up-flows on the observed Dopplershift from a tube with torsional motion. 
Only in the situation that these modes do not produce a significant net velocity component along the line-of-sight, a clear RRE/RBE pair can be detected.

The observation of finding generally more RBEs than RREs suggests that plasma up-flows are present in spicules and result in a bias towards blue-shifts.
For the center-to-limb variation of the detection numbers, we see that the total number of detected events increases towards the limb (i.e., both more detected RBEs and RREs).
In addition, we see that the number of RREs increases relatively more than the number of RBEs. 
Partly, the increase in total number of detected events towards the limb can be explained by the curvature of the Sun and the related increase of the solar surface covered by the fixed size of the FOV when pointing closer to the limb.
In addition, this can be interpreted as resulting from transversal motions gaining weight as the line-of-sight becomes more inclined to the average spicule axis as we go towards the limb.
The net contribution from field-aligned flows would decrease closer to the limb and we see a more balanced ratio of RRE to RBE detections: from RRE/RBE $\approx0.52$ near disk center to $\approx0.74$ near the limb.
% 5.3/10.1 = 0.52   and    14.4/19.4 = 0.74

The transition from RREs into RBEs and vice versa (see Fig.~\ref{fig:long_slit}) can be interpreted as the result of an Alfv{\'e}nic wave associated with the swaying motion.
The exact nature of the wave is difficult to determine since the associated time scale is so short (high propagation speed). 
Only in the high temporal resolution dataset (cadence 0.88~s) we found a number of convincing examples where these changes were occurring in the same physical structure. 
Such high cadence observations could only be acquired at the expense of high spectral resolution so that valuable diagnostic information that could provide more insight into the physical nature of the changes is lacking.

\section{Conclusion}
\label{Sec:conclusion}

In a number of high-quality data sets with different observational characteristics (variation in e.g., spectral line sampled, temporal and spectral resolution, observing angle and target) we find clear evidence of the presence of three different types of motion in the disk counterpart of type II spicules: (1) field aligned flows, (2) swaying motions, and (3) torsional motions.
Our results support the findings of 
\citet{2012ApJ...752L..12D} %De Pontieu et al. 2012 Twist paper 
who established the presence and relative importance of these three kinetic modes in spicules at the limb. 
Rotational motion in spicules had been reported in the earlier literature 
\citep[e.g.,][]{1968SoPh....3..367B, % Beckers review
1968SoPh....5..131P, %Pasachoff+ rotation in spicule - reference from Tsiropoula
1970SoPh...14..310W} %Weart examples of slanted spectral features - reference from Tsiropoula
but only with the aid of modern observational techniques the ubiquity of this type of motion could be determined. 
We also see evidence for propagation of both the swaying and torsional motions, which is compatible with the presence and propagation of Alfv\'enic waves.
Our analysis on the disk counterpart of type II spicules reinforces the conclusions of 
\citet{2012ApJ...752L..12D}: %De Pontieu et al. 2012 Twist paper 
the combined action of all three kinds of motion needs to be considered when interpreting spicule Doppler signals to estimate their contribution to the transport of plasma and energy through the solar atmosphere.

\acknowledgements
%SST
The Swedish 1-m Solar Telescope is operated by the Institute for Solar Physics of the Royal Swedish Academy of Sciences in the Spanish Observatorio del Roque de los Muchachos of the Instituto de Astrof\'{\i}sica de Canarias.  
%B.D.P
B.D.P was supported through NASA contracts NNG09FA40K (IRIS), NNX11AN98G, and NNM12AH46G.  
%Observers
We thank Sven Wedemeyer, Jorrit Leenaarts, Jaime de la Cruz Rodr{\'i}guez, Ada Ortiz Carbonell, Ida Benedikte Hole, H{\aa}kon Skogsrud, and Gregal Vissers for their help during different observation campaigns.
% check names!
% Jaime + Eamon : 2011 May
% Benedikte + Ada : 2011 Sep
% Jaime + Haakon : 2012 May
% Jaime + Gregal + Eamon : 2012 June
% 2008 June: .... Sven, Jorrit, Ada ....
%ADS
This research has made use of NASA's Astrophysical Data Systems.

%\bibliographystyle{aa}
%%\bibliographystyle{apj}
%\bibliography{RRE-paper-1}

\end{document}